\documentstyle[titlepage,fleqn,12pt]{article}
\begin{document}

\begin{center}
{\Large \bf{Langevin dynamics of financial systems: \\
A second-order analysis}} \\
\vspace{2cm}
{\bf Enrique Canessa}\footnote{E-mail: canessae@collaborium.org} \\
{\em The Abdus Salam International Centre for Theoretical Physics,
Trieste, Italy}
\end{center}

\vspace{2cm}
{\baselineskip=20pt
\begin{center}
{\bf Abstract}
\end{center}
We address the issue of stock market fluctuations within Langevin
Dynamics (LD) and the thermodynamics definitions of multifractality
in order to study its second-order characterization given by the
analogous specific heat $C_{q}$, where $q$ is an analogous temperature
relating the moments of the generating partition function for the financial
data signals.  Due to non-linear and additive noise terms within the LD, 
we found that $C_{q}$ can display a shoulder to the right of its main peak
as also found in the S\&P500 historical data which may resemble a classical
phase transition at a critical point.  

\vspace{1cm}
}

\baselineskip=22pt
\parskip=4pt

\pagebreak

\section{Introduction}

Simulated descriptions of at least some universal features of financial
prices has recently been produced by multifractal techniques \cite{Man99}
and also linear \cite{Bou98,Nak98} and non-linear \cite{Ric00} Langevin
Dynamics (LD).  On one hand, multifractals reasonably describe patterns
that resemble sudden (both large and small) market price oscillations,
ensuing bursts of long-range volatility as in the case of stochastic
multi-agent models \cite{Lux99}.  On the other hand, non-linear 
\cite{Ric00} LD generates fat tail distributions of price differences 
as found in the analysis of financial time series \cite{Man95,Sor00}.

There is a vast literature on the characterization of financial data
based on turbulence, multifractality, autocorrelation functions and
power spectra studies (see, {\it e.g.},
\cite{Sch98,Van98,Iva99,Bou00,Kat00}).  However the second-order
derivative relating the increment of price values of assets
$\Delta x(t,1) \equiv x(t+1) - x(t)$ and given by an analogous specific
heat $C_{q}$ similarly to the multifractal studies of diffusion-limited
aggregates (DLA) \cite{Lee88}, has only recently been analysed by the author 
in the case of temporal fluctuations of the S\&P500 stock index \cite{Can00}.
The analogous specific heat is given by the second order derivative of the 
analogous free energy $\tau$ relating the multifractal dimension $D_{q}$ 
({\it c.f.}, Eq.(\ref{eq:tau}) and (\ref{eq:heat}) below).  The $q$-moments 
of the generating partition function for the financial data signals 
$\Delta x(t,1)$ corresponds to an analogous temperature.

As an illustrative example,
the curve (a) in the lower Fig.1 shows the presence of an anomalous
shoulder in $C_{q}$ when considering 3287 (S\&P500) data points including
the largest burst that corresponds to the so-called {\bf Black Monday}
crash measured in October of 1987.  For comparison, in the plotted curves 
(b,c,d) of Fig.1 we remove the crash from the historical data and make the
measurement of $C_{q}$ anew.  Curve (b) is the result of just removing
the single data point for the {\bf Black Monday} ({\it i.e.}, by considering
3286 points), whereas the curve (c) is the result of removing the {\bf
Black Monday} plus all successive data ({\it i.e.}, considering 1970
initial points).  On the other hand, curve (d) is the result of taking 499
initial data points only.
 
Clearly, as one removes or approaches the {\bf Black Monday} data the 
shoulder in $C_{q}$ of curve (a) does not disappear as can be seen in
curves (b,c).  A similar behaviour is observed analysing the data
from few days prior to the S\&P500 crash.  As a difference $C_{q}$ only
displays a typical, single peaked form without anomalies for the data 
prior to, but far away from, the {\bf Black Monday} (as in curve d). 

In this work we make an attempt to characterize the onset of market 
crashes by analysing the second-order derivative relating the increment 
of prices of assets $\Delta x(t,1)$ and given by $C_{q}$.  We shall 
consider the LD of financial systems \cite{Nak98,Ric00} to show that 
the presence of an anomalous shoulder in the specific heats as those 
plotted in Fig.1 can also be a consequence of having non-linear 
stochastic terms within LD.

\section{Linear and Non-linear Langevin Dynamics}

We consider the differential equation with an
additive noise term of the form $g_{t}$ times $\eta_{t}$:
\begin{equation}\label{eq:le}
{dx(t) \over dt} = h_{t}(x(t)) + g_{t}(x(t))\eta_{t} \;\;\; ,
\end{equation}
where as usual $\eta_{t}$ is assumed to be Gaussian-white whose
average and variance are given by $\overline{\eta_{t}}=0$ and
$\overline{\eta_{t}\eta_{t'}}=2D\delta(t-t')$, respectively.

The dynamics of simplest financial systems with random noise
has been proposed in \cite{Nak98,Tak97,Sat98} by setting
\begin{equation}\label{eq:lin}
h_{t}(x(t)) \equiv -\lambda_{t} x(t)   \;\;\;  ;  \;\;\;
g_{t}(x(t)) \equiv 1 \;\;\; ,
\end{equation}
with $\lambda_{t}$ a multiplicative noise that sets the time scale for
deviations from equilibrium.  In the context of economic models,
$x(t)$ may represent price values, $\lambda_{t}$ the rate of change of the
price, and $\eta_{t}$ some external noise of various sources.
In the following we shall refer Eq.(\ref{eq:lin}) as the linear LD.
We add that the scaling behaviour and the $1/f$ noise of trajectories
and velocities with time within linear LD with fluctuating
random forces has been reported in our previous work \cite{Can83}.    

Another simple financial agent model has been introduced in \cite{Ric00}
in analogy with the mean field approach to the Ising model for a magnetic
system characterized by 
\begin{equation}\label{eq:nlin}
h_{t}(x(t)) \equiv J x(t) + b x^{2}(t) -c x^{3}(t)
   \;\;\;  ;  \;\;\;
g_{t}(x(t)) \equiv e + x(t) \;\;\; ,
\end{equation}
with $J$, $b$ and $c$ constants associated with higher-order agent
interactions leading to power laws tails.  The parameter $e$ allows this
agent model to take up the simple Langevin form when $x(t)\rightarrow 0$.
The dynamics of time series from stochastic processes governed by a
similar non-linear LD has also been reported in \cite{Gra00}.

\section{Multifractality and Analogous Specific Heat}

In accordance with the standard of economic notation only relative changes
$\Delta x(t,1)$ are usually relevant.  From these fluctuations, we follow 
\cite{Can00} and construct the generating partition function $Z$ in terms 
of the normalized measures $\mu_{t}>0$, and its moments $q$, by the scaling
\begin{equation}\label{eq:zeta}
Z(q,N) = \sum_{t=1}^{N}\mu_{t}^{q} \sim N^{-\tau(q)} \;\;\; ,
\end{equation}
where
\begin{equation}
\mu_{t} \equiv \frac{|\Delta x(t,1)|}{\sum_{t=1}^{N} |\Delta x(t,1)|}
   \;\;\; .
\end{equation}
In the above we have divided the 1D system of length $L$ into $N$ lines
of length $\ell$ ({\it i.e.}, $N\sim L/\ell$), and have associated this
$N$ with the measured discrete $x(t)$ sequences \cite{Ber98}.

In order to get the thermodynamics interpretation of multifractality
(see, {\em e.g.} \cite{Lee88}), we then consider standard definitions
\begin{equation}\label{eq:tau}
\tau (q) \equiv [q - 1] D_{q} \;\;\; ,
\end{equation}
where $\tau$ represents an analogous
free energy and $D_{q}$ the multifractal dimension.  Of particular interest
here is to consider the analogous specific heat given by the second order
derivative of $\tau$, namely
\begin{equation}\label{eq:heat}
- C_{q}  \equiv \frac{\partial^{2}\tau (q)}{\partial q^{2}} \approx
           \tau(q+1) -2 \tau(q) + \tau(q-1) \;\;\; .
\end{equation}

\section{Discussion}

In the upper curves of Fig.2 we show some typical time evolution of
$\Delta x(t,1)$ governed by the linear LD with $h_{t}$ and $g_{t}$
given in Eq.(\ref{eq:lin}) for different values of
$\eta_{t}=(1-2r_{n'})v$ and $\lambda_{t}=(1 - 2r_{n})u$ with $v=0.1$,
$u=-0.005$, $-0.02$ and $-0.1$, and using the random numbers
$0 \le r_{n,n'}\le 1$.  In the lower curves of Fig.2 we display the
non-linear LD fluctuations of $\Delta x(t,1)$ with $h_{t}$ and $g_{t}$ given
in Eq.(\ref{eq:nlin}) with $J=-1$, $b=0.01$, $c=0.001$, $e=0.005$ and
for different values of $\eta_{t}=(1-2r_{n''})n$ such that $n=1.2$,
$1.5$ and $1.8$ and using random numbers $0 \le r_{n''}\le 1$.

From the curves in Fig.2 it can be seen that $\Delta x(t,1)$ has relatively
small finite values, and exhibits intermittent bursts due to the weak noise
terms.  The presence of peaked narrow fluctuations can increase with both
the multiplicative and additive noise strength.  However, wider volatility
clusters are formed over large periods of time when decreasing, in the linear
LD, the amplitude $u$ of the multiplicative $\lambda_{t}$ noise as
compared to the case of increasing $n$ of the additive noise $\eta_{t}$
within the non-linear LD. It is this difference of behaviour that will lead
$C_{q}$ to display an anomalous shoulder to the right of its main peak to
be shown below.

Let us first see if the generalized dimensions $D_{q}$ for both linear 
and non-linear LD, which are extracted by considering the definition in 
Eq.(\ref{eq:tau}) in conjunction with the scaling of Eq.(\ref{eq:zeta}),
are sufficiently smooth and multifractal for $C_{q}$
of Eq.(\ref{eq:heat}) to be meaningful \cite{Rie95,Pas97}.  
The discrete values of $q$ used range from $-5$ to $+5$ at increments
of $0.033$.

In Fig.3 we display the plots of $D_{q}$ for the different values of $u$ 
and $n$ as used in Fig.2 together with the plots of the limit
$1-D_{q\rightarrow \infty}$ for the linear and non-linear LD
(upper and lower curves, respectively).  If $q<-1$, we observe that
both results for $D_{q}$ follow a typical convergent behaviour according
to multifractal physics.  In the range $u>-0.02$ and $n<1.5$, $D_{q}$ is 
fully multifractal-like.

For lower values of $u<-0.02$ and greater values of $n>1.5$, we find a
non-monotonous decreasing behaviour of $D_{q}$, in correspondence with
the double-peaked form of the respective $C_{q}$ displayed in Fig.4 which
relates the presence of the intermittent bursts in $\Delta x(t,1)$ shown
in Fig.2.  Such a non-monotonous behaviour of $D_{q}$ is more evident in
the case of non-linear LD.  From the $D_{q}$ data for the linear LD case,
the multifractality strength of the $\Delta x(t,T)\equiv x(t+T) - x(t)$ 
sequences at different integer time lags $T \ge 1$, 
{\it i.e}, $1-D_{q\rightarrow \infty}$, does not seem to follow 
a power-law scaling for $1<T<10$ within all $u$ considered. 
As a consequence of the complicated behaviour of the non-linear form of 
Eq.(\ref{eq:nlin}) coupled to $\eta_{t}$, which influences the (thinner) 
volatility clusters shown in Fig.2 (lower curves), it is not 
strightforward that $1-D_{q\rightarrow \infty}$ in this case follows a 
power scaling as in \cite{Pas97}.

Let us see next how this type of $D_{q}$ behaviour given in
Fig.3 influences the analogous specific heat of Eq.(\ref{eq:heat}) and
how it characterizes the temporal large and small intermittent bursts
representing variations in financial prices.  In Fig.4 we plot $C_{q}$
for the LD parameter set used in Figs.2 and 3.  For the linear LD case,
we find that the main peak of our numerical $C_{q}$ curves for
$u>-0.02$ resembles a classical first-order phase transition at a
critical point $q=-1$.  However, the sharp peak turns slightly asymmetric
around the critical point for $u=-0.1$.  Surprisingly, this asymmetry
becomes evident within the non-linear LD.  The analogous specific heat
in this case displays a higher shoulder to the right of the main peak
as a function of $n\ge1.5$ which is in qualitative agreement with the
$C_{q}$ curve for the S\&P500 index data shown in Fig.1 (lower curves).
The presence of the shoulder in $C_{q}$ is then associated to the
non-monotonous behaviour of $D_{q}$ as in Fig.3. 

The same $C_{q}$ shape at a phase transition as in Curve (d) of Fig.1
and Fig.4, has been reported in  the case of DLA \cite{Lee88}.
A non monotonic specific heat shape with a superimposed second peak
(resembling the shoulder of the type of Curve (a) in Fig.1 and
Fig.4 lower curves) has been found in the Hubbard model on a cluster
of magnetic sites \cite{Geo93,Vol97} and, most recently, in a uniform
spin model on a fractal \cite{Les00}.                               

When decreasing $n\le1.2$ in the non-linear LD, the sudden 
intermittent bursts in $\Delta x(t,1)$ as well as the shoulder in $C_{q}$ 
tend to vanish.  It is this latter feature that make us argue that the 
stochastic price behaviour of financial assets may be characterized by 
an analogous $C_{q}$ which resembles the phase transition features as 
measured in multifractal physics.

\section{Conclusion}

We have made a novel attempt to characterize the onset of 
market crashes via an analogous specific heat using the thermodynamics 
definitions of multifractality.  We presented an study of generalized 
dimensions $D_{q}$ of Eq.(\ref{eq:tau}), and the analogous specific heat 
$C_{q}$ of Eq.(\ref{eq:heat}), characterizing the absolute moments of 
the increments of two LD, a linear one ({\it c.f.}, 
Eq.(\ref{eq:lin})) and a non-linear one ({\it c.f.}, Eq.(\ref{eq:nlin})), 
within the context of a multifractal approach to financial time series 
analysis \cite{Can00}.

Tunning parameters in these dynamics introduces wide clusters and 
bursts of volatility in the processes.  This behaviour was shown
to mimics characteristics of S\&P500 data which were measured according to
an analogous phase transition in the behaviour of $C_{q}$.

In this work, the second-order derivative relating the increment of prices
of assets $\Delta x(t,1)$, and given by $C_{q}$, follows similar Legendre
transforms for the thermodynamics analysis of (DLA) \cite{Lee88}.  Our 
findings may relate the results of the kind of those 
described in {\cite{Gra88} where a phase transition-like appears in 
the presence of two distinct phases when analysing the behaviour of $D_{q}$.  
In the latter, these two phases are built out of rare events on the one hand
and some "background" ({\it e.g.}, bursts of volatility in the processes) on
the other hand.  We have found that a shoulder appears in the analogous 
$C_{q}$ due to the presence of wide volatility clusters as found in 
the S\&P500 historical data, or in the non-linear LD.    

The message here is that due to non-linear and additive noise terms 
within the LD, $C_{q}$ can display a shoulder to the right of its main 
peak data as also found in the S\&P500 historical data, which may 
resemble a classical phase transition at a critical point.

\vspace{3cm}

{\it Acknowledgements} \\
The author thanks Dr. V.L.Nguyen for his interest in this work, Prof. 
P. Richmond for providing Ref.\cite{Ric00} in advance and the two 
anonymous Referees for their suggestions.

\newpage
 
\section*{Figure captions}

\begin{itemize}
\item {\bf Fig.1}: Time evolution $\Delta x(t,1)$ of the S\&P500 stock
index for the period 1980-1992 as a function of trading time lags
(in {\it a.u.}) and corresponding analogous specific heat $C_{q}$
obtained using Eq.(\ref{eq:heat}).
The lower curve (a) shows the presence of an anomalous
shoulder in $C_{q}$ when considering 3287 (S\&P500) data points including
the largest burst that corresponds to the crash measured in October of 1987.  
Curve (b) is the result of removing this {\bf Black Monday} crash data point,
whereas curve (c) is the result of removing all successive data.
Curve (d) is the result of taking 499 initial data points only.
 
\item {\bf Fig.2}: Time evolution of $\Delta x(t,1)$ governed by the LD
of Eq.(\ref{eq:le}) using the linear terms in Eq.(\ref{eq:lin})
(upper curves) and non-linear terms of Eq.(\ref{eq:nlin}) (lower curves).

\item {\bf Fig.3}: Generalized dimensions $D_{q}$ and multifractality
strength $1-D_{q\rightarrow \infty}$ for different time langs $T$ for
the linear LD using Eq.(\ref{eq:lin}) (upper curves) and the non-linear
LD using Eq.(\ref{eq:nlin}) (lower curves).  The $1-D_{q\rightarrow \infty}$
curves have been evaluated by considering $\Delta x(t,T) \equiv x(t+T) - x(t)$.

\item {\bf Fig.4}: Analogous specific heat $C_{q}$ of the linear
(upper curve) and non-linear (lower curve) LD derived
using Eq.(\ref{eq:heat}).

\end{itemize}

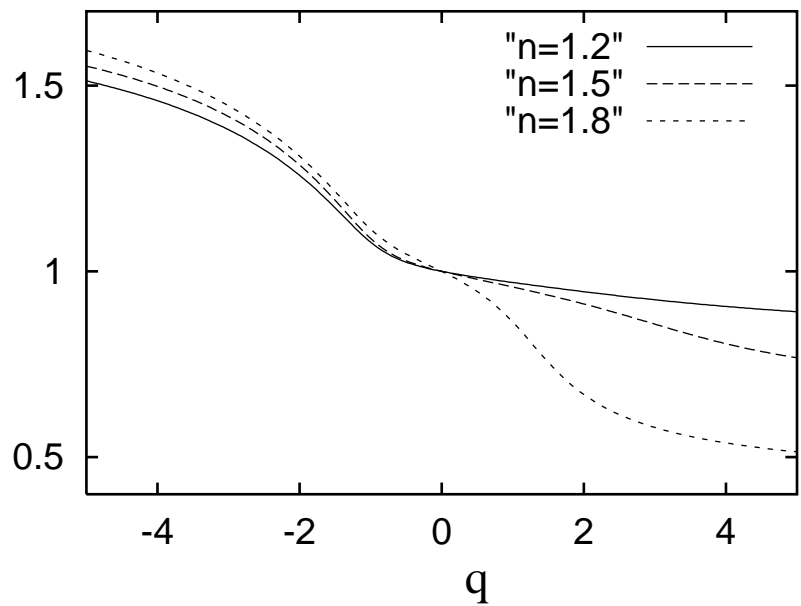
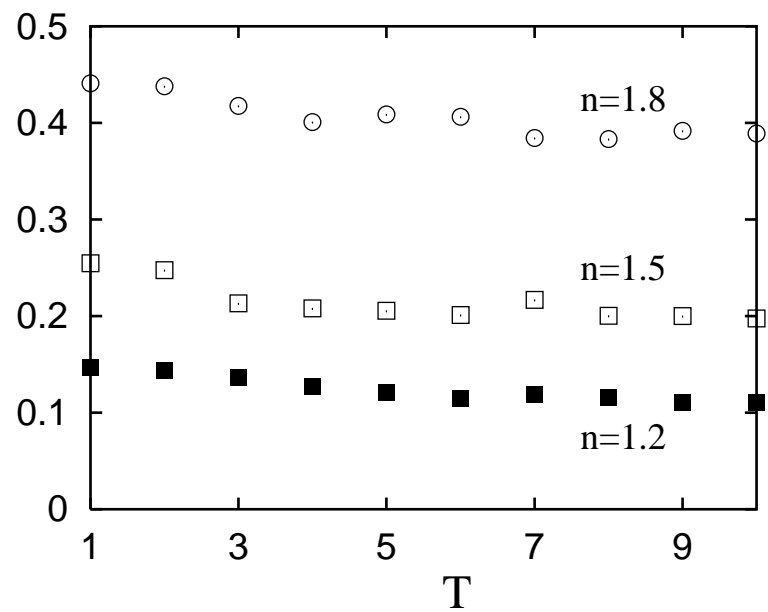
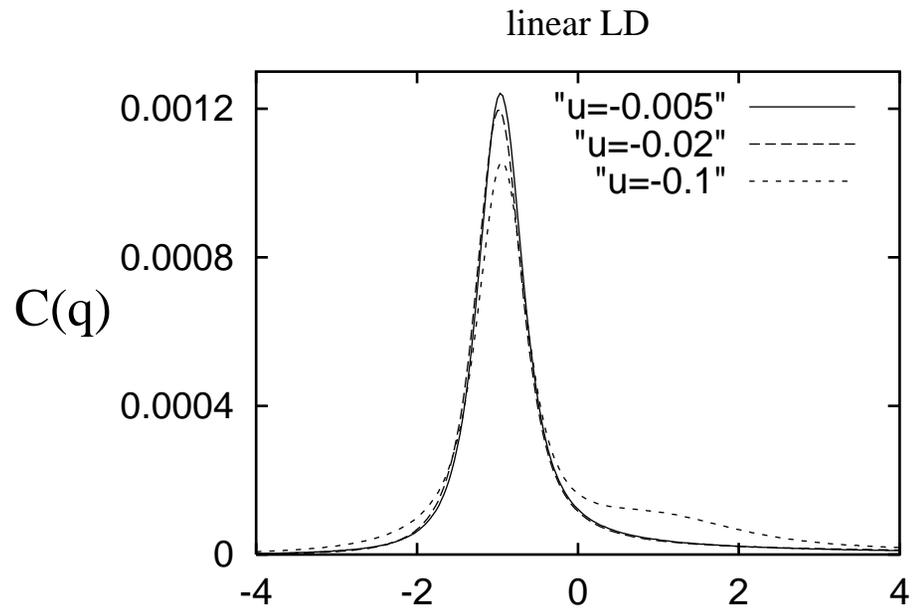
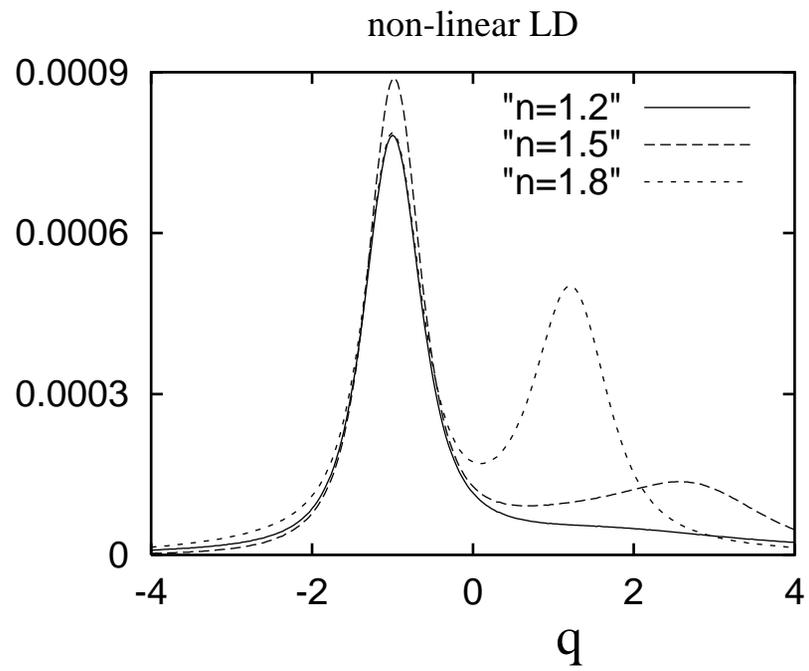
\end{document}